\documentclass{tcibook}
\usepackage{fancyhea}
\usepackage{work}
\usepackage{bm}       
\usepackage{graphicx}
\usepackage{multirow}
\usepackage{lineno}
\usepackage{threeparttable}
\usepackage[normalem]{ulem}
\usepackage{color}
\usepackage[colorlinks = true,
            linkcolor = blue,
            urlcolor  = blue,
            citecolor = blue,
            anchorcolor = blue]{hyperref}
\usepackage{amssymb}

\pdfoutput=1

\def\beq{\begin{equation}}
\def\eeq#1{\label{#1}\end{equation}}
\def\eeqn{\end{equation}}

\newenvironment{Eqnarray}%
   {\arraycolsep 0.14em\begin{eqnarray}}{\end{eqnarray}}
\def\beqa{\begin{Eqnarray}}
\def\eeqa#1{\label{#1}\end{Eqnarray}}
\def\eeqan{\end{Eqnarray}}

\let\bar=\overbar

\def\lsim{\mathrel{\raise.3ex\hbox{$<$\kern-.75em\lower1ex\hbox{$\sim$}}}}
\def\gsim{\mathrel{\raise.3ex\hbox{$>$\kern-.75em\lower1ex\hbox{$\sim$}}}}

\def\del{\partial}
\def\Dslash{\not{\hbox{\kern-4pt $D$}}}
\def\dslash{\not{\hbox{\kern-2pt $\del$}}}
\def\pslash{\not{\hbox{\kern-2pt $p$}}}
\def\ETmiss{\not{\hbox{\kern-4pt $E$}}_T}

\def\Dlr{\mathrel{\raise1.5ex\hbox{$\leftrightarrow$\kern-1em\lower1.5ex\hbox{$D$}}}}

\def\ee{e^+e^-}

\def\MSB{{\bar{M \kern -2pt S}}}
\def\msb{{\bar{\scriptsize M \kern -1pt S}}}

\def\drb{{\bar{\scriptsize D \kern -1pt R}}}

\def\GeV{{\rm GeV}}

\def\authorlist#1#2{
    \vskip 0.4in
\begin{center}\begin{large} {\bf  #1 } \end{large}
    \vskip 0.2in
              #2
     \vskip 0.2in
   \end{center}
}

\begin{document}


\pagenumbering{roman}

\parindent=0pt
\parskip=8pt
\setlength{\evensidemargin}{0pt}
\setlength{\oddsidemargin}{0pt}
\setlength{\marginparsep}{0.0in}
\setlength{\marginparwidth}{0.0in}
\marginparpush=0pt


\pagenumbering{arabic}

\renewcommand{\chapname}{chap:intro_}
\renewcommand{\chapterdir}{.}
\renewcommand{\arraystretch}{1.25}
\addtolength{\arraycolsep}{-3pt}

\setcounter{chapter}{0}


\chapter{Calorimetry}
\label{sec:if06}

\authorlist{A. White, M. Yeh, R. Yohay}
   {(contributors from the community)}

\section{Calorimetry: Executive Summary}

The IF06 Calorimetry group has considered major issues in present and future calorimetry. Input has been taken from a series of talks, group discussions, LOIs, and White Papers. Here we report on two major approaches to calorimeter systems - Particle Flow and Dual Readout, the critical extra dimension of precision timing, and the development of new materials for calorimeters. 

The potential for precision timing at the 10ps level or better opens new possibilities for precise event reconstruction and the reduction of the negative effects of challenging experimental environments. Precise timing can directly benefit calorimetry in several ways ranging from detailed object reconstruction to the mitigation of confusion from pile-up. It can also lead to improved performance for both particle flow and duel readout-based calorimeters.
Given these possible performance enhancements, the focus is now on the study of timing implementation both at the device level and the calorimeter system level. Successful implementation can lead to highly performant calorimeter systems well matched to the demands from both future physics studies and experimental environments.

The construction of future calorimeters matched to the demands of high radiation tolerance, the need for fast timing, and constrained cost puts strong requirements on the properties of active calorimeter materials. Suitable materials should therefore combine high density, fast decay time, and good radiation hardness with good optical quality and high light yield. A range of inorganic scintillators has been developed possessing many of these desired properties. However, use of these materials in future large-scale calorimeter systems demands attention to material costs to assure affordability. Organic based plastic scintillators have seen very wide use, having the advantages of ease of use and relatively low cost. The combination of plastic scintillators with SiPM readout has significantly expanded the flexibility of calorimeter design. Finally, liquid scintillators, with low unit cost, have been widely used where very large volumes are required. An interesting development is the potential of metal doped (water-based) liquid scintillators for use as active materials in sampling calorimeters. 

Particle flow calorimetry makes use of the associations of charged tracks and calorimeter energy deposits to achieve precise reconstruction of hadronic jets and measurement of their energy. Such associations can aslo be effectively used to reduce the effects of pileup. Implementation of a PFA-based calorimeter requires a small cell size and high granularity leading to very high channel counts. Challenges remain in realizing calorimeter systems with up to 100 million cells. Developent is onging for such systems in the areas of power (heat) management, integration of on-board ASICs, and components of signal extraction. A variety of approaches to PFA calorimeters are under developent ranging from scintillator-SiPM systems to a range of gas-based systems using GEMs, Micromegas and RPCs. Also being explored are the potential benefits of precise timing (see above) and moving some elements of a PFA into the front-end electronics.

The dual readout approach to the precise measurement of jet energies makes use of the scintillation light and 
Cerenkov light from showers. The relative amounts of scintillation and Cerenkov light is used to correct the shower energy. The original approach to implementing dual readout calorimetry used a matrix of scintillating fibers and clear optical fibers (for the Cerenkov component) in a block of absorber material. However, with the reduced need for spatial associations (as in particle flow), dual readout can make beneficial use of an electromagnetic front section using homogeneous scintillating crystals which have excellent electromagnetic energy resolution. Development of dual readout systems has followed both of these approaches. In the second case, the separation of the scintillation and Cerenkov signals can be achieved by the use of optical filters and SiPM readout and/or exploitation of the different time structure of the two signals. Development and testing of large-scale systems is needed to demonstrate the feasibility of both the all-fiber and homogeneous electromagnetic section plus fibers approaches to dual readout.

The field of calorimeter research and development remains very active. The precision energy measurement requirements of future physics programs is stimulating innovation in the particle flow and dual readout systems.
The increasing availability of precise timing is adding an important new dimension to system implementation, while the development of a range of fast, radiation-hard active materials is leading to increased flexibility and exciting possibilities for calorimeter system designs.  Future calorimetry at fixed target and colliding beam experiments will be fundamentally multidimensional, providing shower position, time, energy, and a detailed look at shower constituents through the exploitation of an application-specific combination of particle flow techniques, materials with intrinsically good time or energy resolution, and/or dual readout techniques. Sustained R\&D support is needed to achieve this and to move multidimensional calorimetry from prototype to realistic detector.  Scaling to hundreds of thousands or tens of millions of channels while maintaining the required quality is a huge challenge.  Electronics must be developed to allow new features such as fast timing without significantly adding to the power budget or cooling load. It will be necessary to have HEP personnel trained in the new technologies and close tracking of new electronics options emerging in industry.  Finally, effective partnerships with chemists, materials scientists, industries large and small, and radiation facilities must be continued and strengthened to explore the landscape of materials that enable precision calorimetry and to lower the cost.

\section{Precise timing for Calorimetry}
\label{sec:if06_timing}

The primary purpose of electromagnetic and hadronic calorimetry is the measurement of
the energy of charged and neutral particles and overall event energy. However, they are also 
important systems for overall event reconstruction, particle identification and triggering. The
physics goals and the experimental conditions at future colliders require technical advances
in calorimeter technology to fully exploit the physics potential of these facilities. For future
$\ee$ colliders, so-called Higgs Factories, the overall precision of event reconstruction is the
main focus, while future hadron colliders at energies and luminosities significantly beyond
the HL-LHC impose new challenges in terms of the experimental environment.

Precision timing can add an important extra dimension to calorimeter systems. The prospect of
achieving timing resolution at the 10 ps level, the few ps level and even sub-ps level allows
the possibility of a number of new aspects of calorimeter technology implementation and event reconstruction~\cite{Ahmed:2019sim}.
Event reconstruction can benefit from the calorimeter timing capacities at several hierarchical levels: 
timing in cells, in highly granular calorimeters, helps shower reconstruction
and energy corrections, timing of individual showers improves particle identification and
objects reconstruction, and object timing allows event pile-up mitigation and characterization.

Particle identification via time-of-flight systems with a resolution at the 10 ps level allows
pions to be resolved up to about 3 GeV, K-mesons to about 10 GeV, and neutrons
and hyperons to several tens of GeV. While this may be useful for proposed e+e- colliders
detectors, future very high energy hadron colliders will require resolution at the picosecond level.
Precision timing resolution can either be obtained by dedicated timing layers integrated in the 
electromagnetic calorimeter achieving the required resolution for minimum-ionizing particles, 
or by a corresponding resolution for hadronic showers provided by the overall calorimeter system.
Calorimeters with tens-of-picosecond timing can also lead to significant benefits for
reconstruction of heavy long-lived particles which can have distinct signatures.

Particle flow algorithms, using association of tracking and calorimetric information, can also 
benefit from the addition of precise timing information. Since hadronic showers show a complex 
time structure, with late components connected to neutron-induced processes, timing on the cell 
level can have benefits for the spatial reconstruction of hadronic showers. 
A time resolution on the order of a few 100 ps to 1 ns results in a sharper definition of the core
part of the shower, and thus potentially in a better separation of different particles in the
calorimeter, and improved track-cluster assignment in PFA. Full space-time evolution of showers 
could be achieved with 10ps or better timing resolution. Multi-dimensional information from highly 
granular calorimeters including timing can be used in combination with convolutional and graph neural
networks.

The time information can also be used to reconstruct shower shape for longitudinally unsegmented 
fiber sampling calorimeters like the dual-readout calorimeter. This can be done in terms 
of the energy density shape obtained by deconvoluting exponentiating components from detected signals.
Shower length differences can be used to distinguish between electrons and pions.

Precise timing information can also play an important role in event reconstruction with background
suppression, for example in the reconstruction of jet substructure for resolving highly-boosted objects.
For high energy proton collisions, precise timing at the 5-10 ps level will be essential for 
minimizing the effects of very large numbers of pile-up events on separation of signal events in
individual analyses.

Of course, benefiting from all the desirable effects of precise timing described above is dependent
on developing new timing techniques at the sub-10 ps level. An example of a possible solution is
coherent microwave Cherenkov detection, which has demonstrated 2-3 ps timing of electromagnetic 
showers using dielectric-loaded rectangular wave-guide elements.

The actual implementation of timing in calorimeter systems can take a number of forms. Timing information 
can be collected by all active calorimeter cells which, when implemented in a highly granular calorimeter 
enables a full five-dimensional reconstruction of shower activity in the detector, with corresponding 
benefits for pattern recognition, spatial shower reconstruction and separation and energy measurement.
However, the very large number of cells might force a compromise with only a fraction of cells
instrumented for timing. A less challenging and expensive solution could be the use of timing layers,
for instance before and after the electromagnetic calorimeter, and at certain depths in the hadronic
calorimeter.

A variety of possible technologies are being developed for precise timing elements in calorimeters.
These range from Low-Gain Avalanche Detectors and fast silicon sensors to micro-channel plates for
timing layers, and from scintillator/SiPMs or multi-gap resistive plate chambers to small crystal
solutions for volume timing. Given the potential benefits of including fast timing in calorimeter
systems, it is critical that R\&D in the is area should be supported.

\section{Materials for Future Calorimeters}

Future HEP experiments present stringent challenges to calorimeter materials in radiation tolerance, time response and project cost. Here we summarize materials to be developed in the form of inorganic, liquid (oil- and water-based), and plastic scintillators and wavelength shifters to advance HEP calorimetry to face the challenges in radiation hardness, fast timing, and cost-effectiveness. Some of these materials may also find applications for future HEP time-of-flight system, and beyond HEP in nuclear physics, hard X-ray imaging and medical instruments.

Preferred materials for future calorimetry should have high density, good optical quality, high light-yield, fast decay time, good radiation hardness and low cost. High density increases stopping power and reduces calorimeter volume, thus the cost. Good optical transmission enhances signal efficacy and thus uniformity. High light-yield improves signal to noise ratio and thus energy, spatial and timing resolution, and reconstruction efficiency. Fast decay time improves timing resolution and the ability to mitigate high event rate. Good radiation hardness is essential for calorimetry survivability, so is crucial to improve the calorimetry stability. The requirement and development of calorimetric materials are driven by individual experimental goals.

Inorganic scintillators with core valence transition features with its energy gap between the valence band and the uppermost core band less than the fundamental bandgap, allowing an ultrafast decay time. As an example, BaF$_{2}$ (barium fluoride) crystals have an ultrafast cross-luminescence scintillation with 0.5 ns decay time peaked at 220 nm, but also a 600 ns slow decay component peaked at 300 nm with a much higher intensity. The slow component in BaF$_{2}$ crystals causes pileup in a high rate environment but can be suppressed by rare earth doping in crystals. The study of X-ray excited luminescence (XEL) on BaF$_{2}$ with different Y3+ doping levels showed that the light intensity of slow component decreases, while the that of the fast component is maintained, with increasing yttrium doping level. 

The BaF$_{2}$:Y (yttrium doped barium fluoride) crystals have the highest light yield in the 1st nanosecond and the highest ratio between the light yield in the 1st nanosecond and the total light yield, which are the figures of merit for a TOF system. Future HEP experiments at the energy and intensity frontiers require ultrafast calorimetry to mitigate high event rate and break the picosecond timing barrier, where BaF$_{2}$:Y crystals and solar-blind VUV photodetectors are under development for the proposed Mu2e-II experiment. An ultrafast BaF$_{2}$:Y total absorption calorimeter is also considered by the RADiCAL consortium.

Scintillating glass (SciGlass) is a new type of active material that is being engineered, in the context of calorimeter R\&D for the Electron-Ion Collider (EIC), to have similar performance and radiation hardness to lead tungstate at a fraction of the cost.  Development has benefited from an SBIR/STTR grant shared between the Vitreous State Laboratory of the Catholic University of America and Scintilex, LLC.  Since early 2019, the SciGlass prototypes have grown from coin-sized samples to 40-cm bars representing 20 radiation lengths.  SciGlass is radiation hard to 1000 Gy and 10$^{15}$ n/cm$^{2}$, has a response time of 20-50 ns, and has transmittance of $\sim$80\% at the emission peak of $\sim$440 nm~\cite{AbdulKhalek:2021gbh}.  Unlike lead tungstate, SciGlass is stable with temperature.  Work continues to scale up production to the sizes needed for the EIC electromagnetic calorimeter in a cost-effective way.

\sloppy High radiation doses are expected at the forward calorimeter for future collider experiments. Lu$_{2(1-x)}$Y$_{2x}$SiO$_{5}$ (Lutetium Yttrium Orthosilicate) or LYSO:Ce (Cerium-doped Lutetium Yttrium Orthosilicate) and Lu$_{3}$Al$_{5}$O$_{12}$ (Lutetium Aluminum Garnet) or LuAG:Ce (Cerium-doped Lutetium Aluminum Garnet) show high stopping power, high light output, fast decay time and excellent radiation hardness against ionization doses and hadrons. LYSO:Ce crystals are used to construct a barrel timing layer (BTL) for the CMS upgrade for the HL-LHC, where the attenuation length of scintillation light is required to be longer than 3m after radiation doses of 5 Mrad from ionizations of 2.5$\times10^{13}$
charged hadrons per $cm^{2}$ and 3$\times10^{14}$ 1-MeV equivalent neutrons/$cm^2$. While LYSO:Ce crystals satisfy such requirement, LuAG:Ce ceramics shows a factor of two better radiation hardness than LYSO:Ce crystals. They both were proposed by the RADiCAL consortium for the HL-LHC and the proposed FCC-hh in development of an ultra-compact, ultra-radiation hard and longitudinally segmented shashlik calorimeter.  Along with the development of intrinsically radiation hard materials, more R\&D is needed to understand damage recovery, for example by exposure to light~\cite{Sahbaz:2022nit}, to provide more options for future calorimeter designs that will surely be constrained by cost and mechanics.

The cost of inorganic crystals (\$1s per cc) is one of the major challenges for large calorimetry experiments. The proposed lepton Higgs factory requires good EM and jet resolutions. The dual readout CalVision crystal ECAL followed by the IDEA fiber HCAL provides an excellent option. Because of its total absorption nature, the HHCAL concept promises the best jet mass resolution, however, requires deployment of inorganic crystals at large volume. Crucial R\&D is to develop a cost-effective approach by either reducing manufacture material cost or enhancing stopping power of heavy inorganic scintillators to improve the project affordability. 

WLS (wavelength shifter) capillaries and fiberoptic filaments form the optical “bridges” that connect the scintillation light emission from the scintillator plates to photosensors. The light collection efficiency of a calorimeter thus depends on light propagation (absorption and reemission) between the scintillator and the WLS and their energy response to the quantum efficiency of the photosensors. Ideally if sufficiently rad hard, photosensors could be mounted directly to scintillation plates positioned at shower max to avoid the use of WLS. Such a configuration could greatly improve precision timing measurement and light collection efficacy, and reduce installation cost. The performance of mixed modular configurations using different inorganic scintillators and photosensors  with or without WLS is under investigation by RADiCAL.

Organic-based scintillator calorimeters, including plastics, pure liquid scintillator (LS), and water-based liquid scintillator (WbLS), have fast pulse response (1-2 ns) with adequate light-yield (10$^4$ photons/MeV). Plastic scintillators are widely used for ionizing radiation detection due to modest cost (\$10s per kg) and scale-up availability. The most common type of plastic scintillator is composed of selected fluors and wavelength shifters in a plastic base that is an aromatic compound with delocalized $\pi$-electrons. A detector configuration of plastic scintillators tiled with WLS fibers is a popular choice for sampling calorimeters. Most plastic scintillators employ polyvinyltoluene (PVT) and polystyrene (PS) as the base resins for fabrication. A new thermoplastics acrylic scintillator aiming to load scintillator materials and high-Z elements directly into acrylic monomers is under development. The mechanically advantageous property, good thermal resistance, and reduced aging effect make the acrylic a promising candidate for constructing large plastic detectors. By using modern SiPMs, a multilayer acrylic detector could further provide excellent position and energy reconstruction. The success of this improved acrylic-based scintillator material has applications in calorimetry and other particle physics fields, such as neutrino and dark matters. 

Liquid scintillator and water-based liquid scintillator are most cost-effective (\$1s per kg) and less sensitive to radiation damage with long optical transparency (attenuation length of $\geq$10m at 450nm) that are capable of deployment in conjunction with any detector compartments in high dose environments. The chemical safety and compatibility have been largely improved over the past decades. The modern liquid scintillators are not flammable and can be deployed in most plastic and/or steel materials. With 30\% mass loading of heavy elements (i.e. tungsten) in a liquid scintillator, a radiation length of less than 10cm can be achieved. The fabrication and stability of metal-doped liquid scintillators (i.e., gadolinium, lithium, boron) at ton-scale have been demonstrated by the Daya Bay, LZ and PROSPECT experiments. The principal of loading high-Z elements at high mass percent (i.e., lead, indium, tellurium at $\geq$10\%) was proved by solar (LENS) and double-beta decay (SNO+) scintillator research. The application of WbLS in nonproliferation and multi-physics detection is in progress by WACTHMAN and THEIA. Various large-scale testbeds (BNL 1- and 30-Ton Demonstrators and LBL 3-Ton EOS) are under construction to prove the WbLS deployment feasibility at kiloton scale. Significant progress towards developing extractants and techniques that will allow the loading of high-Z elements into high flashpoint and less chemical-aggressive scintillator materials has been achieved by different frontiers. Promising scintillator materials identified for calorimetry are linear alkylbenzene (LAB), cyclohexylbenzene (PCH), 1-phenyl-1-xylyl-ethane (PXE) and di-isopropylnaphthalene (DIN). The approach here is to extend the usage of metal-doped (water-based) liquid scintillators, built on the advanced techniques developed from either deployed or developing neutrino and dark matter search, as active materials for sampling calorimeters through measurements of light collection efficiency, uniformity, and radiation hardness at test beams. 

\section{Particle flow calorimetry}

Particle flow (PF) calorimetry is an experimental technique to realize ultra-precise measurements of hadronic jets and thus enable discrimination between $W$, $Z$, and $H$ bosons reconstructed in multijet final states.  This is a requirement for achieving sub-percent precision on measurements of the Higgs boson mass, total width, and couplings at a future $\ee$ collider.  Although these measurements were the original motivation for PF calorimeter R\&D, the utility of the PF design in assigning showers to the correct event vertex in high-multiplicity environments has now also been recognized.

PF reconstruction leverages the efficient association of tracks and calorimeter deposits afforded by finely segmented detectors to make hypotheses about the exact particle content of an event (i.e. the 4-vectors of each charged hadron, neutral hadron, photon, electron, and muon).  In jet reconstruction, high resolution tracking information can then be used to estimate the momentum of charged candidates over a large momentum range, leaving the momentum assignment to the relatively low resolution calorimeter information for only neutral candidates and very high momentum candidates.  By clustering particle candidates, each of which has been assigned its momentum by the ``best'' detector, large gains in jet energy resolution can be realized over non-PF algorithms that cluster energy deposits alone, given that the charged component of a typical jet is 62\%~\cite{THOMSON200925}.  More effective association of calorimeter hits to tracks with associated primary or secondary vertices also helps to reduce pileup contamination in light and heavy flavor jets.

Benchmarks for PF calorimeter performance have been set by the requirements of precision Higgs and electroweak physics at proposed $\ee$ facilities.  Broadly speaking, jet energy resolutions better than 5\% are expected for energies between 50 and 250 GeV, with a stochastic term of around 30\%/$\sqrt{E\,[\GeV]}$, where $E$ is the jet energy.  This ensures a di-jet mass resolution of approximately 2.7\%, comparable to the total widths of $W$ and $Z$ bosons, that facilitates clean separation of $W$ and $Z$ hadronic decays.  A measurement of the Higgs total width in $ZH$ events using the recoil method, where the initial-state $\ee$ kinematics are used to constrain the final-state Higgs 4-vector from the measured recoiling $Z$, demands efficient $Z\rightarrow q\bar{q}$ identification ($q$ is any flavor of quark) independent of the number of additional jets in the event.  The engineering necessary to hit these benchmarks should, additionally, allow for efficient reconstruction of hadronic tau decays and long-lived particles decaying to SM particles within the detector volume enclosed by the calorimeter.

Although PF calorimetry has undergone extensive R\&D, a number of challenges still need to be overcome to construct a realistic collider detector.  Among the most important is the design of assembly, quality assurance, and quality control processes that scale to $10^{7}-10^{8}$ channels~\cite{Sefkow:2015hna} at a reasonable cost.  Ref.~\cite{Ruchti:2022ixx} states that ``[e]ven though the overall channel counts of PF calorimeters are hundred-thousands to millions and the covered areas in the active layers are hundreds of square meters, these products are a niche for most industry areas.  Usually, only the components and some of the first assembly steps can be bought from or done by companies.  Examples are silicon sensors (but the requirements are unusual, so there are very few suppliers), SiPMs, Micromegas, and production and component assembly of PCBs.  The production of modules (joining the active sensor with the readout electronics) and everything from there on is typically not possible in industry.''  Not surprisingly, efficient simulation of the enormous channel counts in a PF calorimeter is time-consuming.  Simulation is critical to detector design and eventually data analysis.

A related challenge is heat management and thermal performance when trying to fit a greatly increased number of channels into roughly the same physical volume as existing energy frontier detectors.  Complex front-end ASICs that perform first-level data reduction are typically embedded into the active layers in PF calorimeter designs, as there is no room to bring all of the raw data signals off detector via optical or electrical cables.  Power management of these ASICs is crucial to fit within the overall cooling capacity of the experiment.  In high luminosity or high radiation applications using silicon sensors or SiPMs, the thermal requirements of the active media are significant to maintain low leakage currents (noise).  In addition to minimized power consumption, front-end electronics, power converters, cables, and connectors need to have a tiny vertical height to minimize the air gap between sampling layers.  An increased air gap increases the effective Moli\`{e}re radius of the electromagnetic section of the calorimeter, which may compromise the track separation needed to fully profit from the PF approach.  In low-luminosity $\ee$ scenarios, gaps as small as 1 mm are feasible, but this has not yet been demonstrated on a large scale, or for high-luminosity prototype designs where the data volume and heat load of the front-end electronics is more significant.

Finally, in a muon collider or hadron collider detector, PF front ends need to withstand significant ionizing and non-ionizing radiation doses.  For example, at the HL-LHC, the calorimeters must survive up to 1 MGy of ionizing radiation and $10^{16}\,\mathrm{n}_{\mathrm{eq}}/\mathrm{cm}^{2}$ of non-ionizing fluence.  Although this constraint is not unique to PF calorimeters, it places extra pressure on the embedded front ends that are typical of these calorimeters.  In silicon-based front end designs, for which radiation-induced leakage current growth can be mitigated with cooling, complicated cooling systems need to be deployed.

There is a broad landscape of PF R\&D that has paved the way for wide acceptance of this technology and interest in deploying it at scale to solve future HEP calorimetry challenges.  PF concepts are integral to detector designs for the HL-LHC, ILC, FCC-ee, FCC-hh, CEPC, and CLIC.  Table~\ref{tab:PF-summary}, reprinted from Ref.~\cite{Ruchti:2022ixx}, summarizes the current proposals.  All of the concepts in Table~\ref{tab:PF-summary} have been constructed as small-scale prototypes and subjected to beam tests.  Large-scale technological prototypes, demonstrating the feasibility of both assembly and operation of a full-size calorimeter, have been constructed for the SiW ECAL (CALICE), FoCal (ALICE), scintillator ECAL (CALICE), analog HCAL (CALICE), and digital HCAL (CALICE).  Of these, the SiW ECAL, FoCal, AHCAL, and DHCAL have undergone beam testing.  One effort that has transcended R\&D and is shortly entering full production is the HGCAL upgrade for the CMS detector, consisting of both silicon and scintillator active media.



\begin{table}[htbp]
\caption{Overview of the characteristics of several particle flow calorimeter concepts and technologies.  Reprinted from Ref.~\cite{Ruchti:2022ixx}.}
\begin{minipage}{\textwidth}
\begin{center}
\begin{tabular}{|l|l|l|l|l|l|l|}
\hline
name & purpose & project & \begin{tabular}{@{}l@{}}active\\material\end{tabular}  & channel size & readout & \begin{tabular}{@{}l@{}}\# of layers\\(depth)\end{tabular} \\
\hline
CALICE SiW ECAL & ECAL & ILC\footnote{also for CLIC \& FCC-ee} & silicon & 5 $\times$ 5 mm$^{2}$ & analog & 30 (24$X_{0}$) \\
SiD ECAL & ECAL & ILC & silicon & 13 mm$^{2}$ & analog & 30 (26$X_{0}$) \\
HGCAL Si & ECAL\footnote{silicon also used in HCAL part} & HL-LHC & silicon & 52-118 mm$^{2}$ & analog & 28 (25$X_{0}$) \\
FoCal & ECAL & HL-LHC & silicon & 30 $\times$ 30 $\mu$m$^{2}$ & digital & 28 (25$X_{0}$) \\
CALICE Sci-ECAL & ECAL & ILC\footnote{also for CEPC} & SiPM-on-tile & 5 $\times$ 5 mm$^{2}$\footnote{effective size, strips have 5 $\times$ 45 mm$^{2}$} & analog & 30 (24$X_{0}$) \\
RADiCAL & ECAL & FCC-hh & \begin{tabular}{@{}l@{}}crystal +\\WLS\footnote{wavelength-shifting fiber}\end{tabular} & 4 $\times$ 4 mm$^{2}$\footnote{effective size at shower max; module cross-section is 14 $\times$ 14 mm$^{2}$} & analog & 29 (25$X_{0}$) \\
\hline
CALICE AHCAL & HCAL & ILC\footnote{also for CEPC, CLIC \& FCC-ee} & SiPM-on-tile & 3 $\times$ 3 cm$^{2}$ & analog & 40 (4$\lambda_{I}$) \\
HGCAL Scint & HCAL & HL-LHC & SiPM-on-tile & 6-30 cm$^{2}$ & analog & 22 (7.8$\lambda_{I}$)\footnote{contains also pure silicon and mixed layers} \\
CALICE DHCAL & HCAL & ILC & RPC & 1 $\times$ 1 cm$^{2}$ & digital & 40 (4$\lambda_{I}$) \\
CALICE SDHCAL & HCAL & ILC & RPC & 1 $\times$ 1 cm$^{2}$ & semi-digital & 40 (4$\lambda_{I}$) \\
\hline
\end{tabular}
\end{center}
\end{minipage}
\label{tab:PF-summary}
\end{table}

The use of silicon active layers is typically restricted to electromagnetic sections, high-$\eta$ forward detectors, and environments with high pileup or non-ionizing fluence.  In these cases, high lateral granularity is necessary to distinguish nearby tracks, and the advantage of silicon is the ability to pattern small cell sizes of hundreds of square microns to hundreds of square millimeters reliably in industry.  An example is the development of MAPS electromagnetic calorimetry for
the SiD detector~\cite{instruments6040051}. This technique has shown the potential for excellent photon reconstruction and separation, and excellent pi-zero reconstruction while delivering very good electromagnetic energy resolution. The addition of timing layers would also be a natural extension of the MAPS approach.  Silicon readouts are usually analog, which helps to determine the shower shape and energy deposition more precisely, but in the ALICE FoCal upgrade for $3.4 < |\eta| < 5.8$, digital readout with small pixels is sufficient.  The CALICE scintillator ECAL achieves an effective cell size of 5 $\times$ 5 mm$^{2}$ from 45-mm-long scintillator strips oriented perpendicular to each other in alternating layers, with the strips coupled to silicon photomultipliers (SiPMs).  This is an example of how questions of cost control and scalability are already being addressed at the R\&D level.

SiPMs are a key enabling technology for PF calorimetry.  They replace bulky photomultiplier tubes, making feasible the use of scintillating media in compact, highly granular designs.  Scintillators may be preferred in PF designs due to their low cost (plastics, for example, are used in the HGCAL hadronic section and CALICE AHCAL) or utility in specific challenging applications (the FCC-hh RADiCAL use case).  In the latter case, SiPMs are coupled to wavelength-shifting fibers that carry light from multiple crystal scintillator layers to the rear of the detector module.  In the former case, each channel (scintillator tile) is read out by an individual SiPM placed right underneath the tile in the active layer (``SiPM-on-tile'').

In digital and semi-digital (2-bit) readout hadronic calorimeter designs, small cell sizes are implemented in a cost-effective way using resistive plate chambers.  More advanced gaseous detectors, such as GEMs, Micromegas, and MPGDs may also be used to form the cells.

The HEP community has demonstrated its strong interest in pursuing this technology in multiple applications in various stages of development.  These include the HL-LHC (CMS HGCAL), $\ee$ collider alternatives (SiD and ILD for the ILC, the CLIC detector concept, CLD for the FCC-ee, and the baseline CEPC detector concept), FCC-hh, and a possible muon collider.  Among these, the CMS HGCAL upgrade for the HL-LHC is at the most advanced stage of development, with delivery of the production silicon sensors slated to begin in January 2023.  The HGCAL will be a crucial demonstration of the particle flow concept.  Lessons learned from the assembly, operation, and PF reconstruction of the HGCAL will influence simulation techniques, design, and costing for future PF calorimeters, and will reveal the areas in which R\&D is most needed.

PF calorimetry development has synergies with developments in other areas of HEP instrumentation.  Silicon detectors, SiPMs, fast scintillators, and gas ionization detectors are all good candidate active media, depending on the use case.  Smart, low-power, radiation-tolerant front end electronics are needed to realize compact designs.  PF reconstruction provides a benchmark for optimizing novel computational methods, like the use of machine learning in triggering or particle reconstruction.  The computational problem of particle flow reconstruction, in which charged and neutral hadrons, electrons, muons, hadronic tau decays, photons, and jets must be identified from associations between tracks and highly granular calorimeter deposits, may be particularly well suited to machine learning methods as opposed to a traditional deterministic algorithm~\cite{Pata:2021oez}.  In turn, PF reconstruction requires performance advances that can speed up detector simulation or improve jet energy resolution.  Breakthroughs in any of these related technologies can drive significant progress in PF calorimetry, be it cost reduction, simplification of engineering, increased radiation hardness, or improved physics performance.

To realize the full potential of PF calorimetry, and to ensure that it is ``shovel-ready'' when the next large experiment is approved, R\&D is needed to solve outstanding problems.  Extensive beam testing, especially of large-scale technological prototypes, is needed to study integration issues and how they affect electromagnetic and hadronic energy resolution.  As stated above, more research is needed in front end design, especially for high occupancy and high radiation applications like a future hadron or muon collider.  Finally, the challenge of rethinking traditional assembly and QA/QC procedures to find solutions that scale to tens to hundreds of millions of channels cannot be overlooked.  There is a growing need to incorporate lessons from product and process engineering into instrumentation labs, either through personnel decisions or by adapting the training of traditional physics graduate students and postdocs.  Similarly, academic-industrial partnerships need to be formed with this scale in mind.

With a holistically designed PF calorimeter and tracker, unprecedented jet energy resolutions can be realized.  This enables $W$/$Z$/$H$ separation in hadronic final states and a measurement of the Higgs total width at an $\ee$ machine.  The detailed jet substructure imaging made possible by high granularity in PF designs can also greatly improve long-lived particle, tau lepton, and boosted object reconstruction.  For these reasons, particle flow has emerged as a leading candidate design for future collider detectors.  With targeted R\&D, the scientific promise of particle flow calorimetry can be exploited in the medium term.

\section{Dual readout calorimetry}

Ref.~\cite{Pezzotti:2022ndj} introduces dual-readout (DRO) calorimetry with the following:

\begin{quote}
    Dual-readout calorimetry is a proven technique for improving calorimeter resolutions and yet its full potential remains to be explored.  As inelastic collisions produced in a hadronic shower are associated with a lower response due to energy lost to binding energies, neutrons migrating far from the shower, neutrinos produced in pion decays, and other sources, the method uses proxies to estimate their number.  Since fluctuations in the average response due to fluctuations in the number of inelastic collisions in the shower dominates the hadronic resolution, it can be greatly improved using an energy scale correction based on this proxy.  In the classic method, the total energy of all ionizing particles is estimated via scintillation light, and the energy depositions of protons produced in inelastic collisions (often via neutron interactions) are estimated via the fraction of the shower particles with a velocity too low to produce $\check{\mathrm{C}}$erenkov light.  Current state-of-the-art calorimeter resolutions are the result of pioneering work by the DREAM/RD52 and IDEA collaborations.
\end{quote}

DRO calorimetry is largely motivated by the same physics as PF calorimetry: precise measurement of jets in $ZH$ production at an $\ee$ collider.  Both methods use additional information about the hadronic shower, beyond simply its energy, to improve jet energy resolution.  In the PF approach, association to tracks allows the charged component of the shower to be precisely estimated and measured by a high-resolution tracker, leaving only the neutral component to be measured by the calorimeter.  In the DRO approach, the electromagnetic component of the shower is estimated from $\check{\mathrm{C}}$erenkov light, and the relative amounts of $\check{\mathrm{C}}$erenkov and scintillation light are used to correct the energy of each shower according to its measured $e/h$ ratio.  PF and DRO methods target the same benchmark of 30\%/$\sqrt{E\,[\GeV]}$ in the stochastic term of the jet energy resolution.

PF calorimeters designed with $ZH$ physics in mind typically have EM energy resolutions of about 15\%/$\sqrt{E\,[\GeV]}$, which is sufficient for jet reconstruction and $H\rightarrow\gamma\gamma$ measurements but not state-of-the-art.  EM resolution is limited by the effective Moli\`{e}re radius that can be achieved due to the finite air gap for readout electronics.  However, a number of physics measurements that profit from a large-statistics ``$Z$-pole'' run of a future $\ee$ collider, including heavy flavor physics with neutral pions, searches for flavor-violating decays such as $\tau\rightarrow\mu\gamma$, or searches for non-universal couplings of the neutrino species via $\ee\rightarrow\nu_{e}\bar{\nu}_{e}\gamma$, could be greatly improved by the addition of state-of-the-art EM energy resolutions.  Because DRO does not rely as heavily on longitudinal granularity to perform spatial associations, it can be integrated with homogeneous scintillating crystals (with dimensions of order $1 \times 1 \times 20$ cm$^{3}$) that have excellent intrinsic EM energy resolution.

Most R\&D for realizing a DRO hadronic calorimeter has focused on a ``spaghetti'' geometry, where scintillating and clear (for $\check{\mathrm{C}}$erenkov light production) optical fibers are installed in a prism (tower) of passive absorber material such that the fibers are parallel to the longest (projective) dimension of the prism.  Towers are typically 2-2.5 m long.  The sampling fraction is related to the pitch between fibers (2-4 mm) and the fiber diameter ($\sim1$ mm).  Practically speaking, such a device is assembled by threading each fiber through a copper or brass capillary and densely packing the capillaries into an enclosing tower module.

Scintillating crystals under study for homogeneous EM calorimeters also produce $\check{\mathrm{C}}$erenkov light, although a dedicated radiator could be added to the crystal matrix to boost the yield.  DRO could be added to a homogeneous EM crystal calorimeter in one of two ways: usage of two separate filter+SiPM assemblies at the rear of the crystal, each sensitive to either $\check{\mathrm{C}}$erenkov light or scintillation light; or exploitation (in additional hardware or in software) of the different time structures of promptly emitted $\check{\mathrm{C}}$erenkov light and more slowly emitted scintillation light.

Research into the design and performance of a DRO calorimeter for an $\ee$ collider has proceeded along multiple fronts.  One of the detector concepts for either FCC-ee or CEPC, the Innovative Detector for an Electron-positron Accelerator (IDEA), utilizes a DRO ``spaghetti'' calorimeter like that described above, based on the SPACAL and RD52 calorimeters.  Simulations suggest that a jet energy resolution of 38(30)\%/$\sqrt{E\,[\GeV]}$ is achievable with 2(2.5)-m-long towers.  Developments of the homogeneous crystal ECAL with DRO have been made possible recently by the availability of SiPMs with good sensitivity in the red portion of the spectrum.  Even though $\check{\mathrm{C}}$erenkov light is enhanced at shorter wavelengths, most of the short-wavelength light is reabsorbed by the crystal as it travels to the SiPM, leaving mostly longer-wavelength $\check{\mathrm{C}}$erenkov light for detection.  This line of R\&D has led to the SCEPCAL concept of a front crystal EM section followed by a rear spaghetti hadronic section.  Simulated hadronic (not jet) energy resolutions are around 27\%/$\sqrt{E\,[\GeV]}$ for the combined ECAL+HCAL system.  The EM resolution is about 3\%/$\sqrt{E\,[\GeV]} \bigoplus 0.5$\%, typical of crystal calorimeters, and obviously much better than what is achievable in sampling PF or DRO designs.

Significant R\&D is still needed to fully realize the potential of DRO.  Large-scale prototypes of the IDEA and SCEPCAL calorimeters need to be constructed to understand which parts of the engineering are cost drivers, which need further R\&D, and how a realistic system will perform.  Such efforts have already begun for the IDEA fiber calorimeter, considering a stacked capillary arrangement as described above as well as a 3D-printed copper alveolar tower into which fibers are inserted.  On the SCEPCAL front, a key open question is how to ensure that readout of the red end of the $\check{\mathrm{C}}$erenkov spectrum will yield a large enough signal with which to perform the DRO correction.  Due to the large size of collider detectors, another important question is how to drive down the cost of high-performing crystal scintillators.  For this, systematic surveys of many candidate materials, both known and novel, should be performed.  In the long term, research into new optical materials, such as photonic crystal fibers for increased $\check{\mathrm{C}}$erenkov light collection, quantum dot and semiconductor nanoparticle wavelength shifters, and fibers that maintain polarization information, can improve the performance and/or reduce the cost of DRO fiber calorimeters.

Readout of each fiber by a SiPM opens the door to jet imaging and position reconstruction using algorithmic or deep learning techniques.  This can enable gluon/jet discrimination, studies of jet substructure, or identification of tau leptons or boosted objects.  Indeed, the extra front end information that DRO relies on requires advances in electronics for efficient processing and reduction.  Areas of R\&D in this direction include system-on-chip waveform digitizers with real-time analysis, field programmable analog arrays on the front end, and digital SiPMs.  With flexibly designed readouts, it may also be possible to use the time structure of the $\check{\mathrm{C}}$erenkov and scintillation signals to estimate the shower EM fraction online and improve triggering.

A novel idea for smaller systems for forward calorimetry, high intensity experiments, and orbiting systems is the use of photomultiplier tubes (PMTs) for the direct detection of shower particles.  The PMTs act as direct calorimeter sensors to detect shower particles via $\check{\mathrm{C}}$erenkov light in the PMT window, and/or by direct secondary emission from shower particles traversing the dynodes.  The secondary emission proportional to $dE/dx$ provides compensating information.

There is no reason that the advantages of PF and DRO cannot be combined in certain applications to yield even more flexibility, and in fact some proposals seek to do just that.  The REDTOP~\cite{REDTOP} experiment at Fermilab would produce some $10^{13}$ $\eta$ particles to serve as a laboratory for studying fundamental symmetry violations, carrying out searches for rare new physics scenarios, and studying SM physics at medium energies with exquisite precision.  Its calorimeter, ADRIANO2~\cite{Gatto:2022abc}, consists of a sandwich of lead glass and plastic scintillator layers.  Each layer is subdivided into optically isolated tiles that are each read out by an individual SiPM.  The lead glass tiles are sensitive to the $\check{\mathrm{C}}$erenkov component, while the plastic tiles are sensitive to the scintillation component.  This configuration, which is currently being studied as part of a multi-year R\&D and test beam campaign, can provide the excellent energy resolution, position resolution, and particle ID needed to discriminate prompt photons from neutrons and $\pi^{0}$s in a high multiplicity event.

DRO is a promising direction in calorimetry that allows for shower-by-shower $e/h$ compensation, ultimately yielding ultra-precise jet energy measurements.  The RD52/DREAM collaboration has established the basics of DRO, but multiple R\&D questions remain unanswered.  This R\&D has been taken up by groups pursuing the IDEA detector or the SCEPCAL calorimeter for a future $\ee$ machine.  It is crucial to demonstrate the feasibility of a large-scale installation, while also pursuing ``blue-sky'' research into novel materials that can provide excellent EM performance and a strong $\check{\mathrm{C}}$erenkov signal.  Such a program will advance DRO into a mature technology for future collider detectors.
  




\bibliographystyle{JHEP}
\bibliography{Instrumentation/IF06/myreferences} 












\end{document}